\documentclass[prl,superscriptaddress,twocolumn,
showpacs, preprintnumbers,
amsfonts,amsmath,amssymb]{revtex4}

\usepackage{epsf,psfrag}
\usepackage{graphicx}
\usepackage{dcolumn}
\usepackage{bm}

\catcode`\@=11
\def \be  {\begin{equation}}
\def \ee  {\end{equation}}
\def \ba  {\begin{eqnarray}}
\def \ea  {\end{eqnarray}}
\def \baa {\begin{eqnarray*}}
\def \eaa {\end{eqnarray*}}
\def \bb  {\begin {thebibliography} }
\def \eb  {\end{thebibliography}}
\def \lab #1 {\label{#1}}

\def \matrix #1 {\left(\begin{array}{cc} #1 \end{array}\right)}

\def \tr {\mathop{\rm tr}\nolimits}

\newcommand{\as}{\ifmmode\alpha_{\rm s}\else{$\alpha_{\rm s}$}\fi}
\newcommand{\asbar}{\ifmmode\bar{\alpha}_{\rm s}\else{$\bar{\alpha}_{\rm s}$}\fi}

\font\cmss=cmss12 
\def\inbar{\,\vrule height1.5ex width.4pt depth0pt}
\def\IC{\relax\hbox{$\inbar\kern-.3em{\rm C}$}}
\def\IZ{\relax{\hbox{\cmss Z\kern-.4em Z}}}
\def\IR{{\hbox{{\rm I}\kern-.2em\hbox{\rm R}}}}

\def\IP{{\hbox{{\rm I}\kern-.2em\hbox{\rm P}}}}
\def\II{\hbox{{1}\kern-.25em\hbox{l}}}

\newbox\lett\newdimen\lheight\newdimen\lwidth
\def\ontop#1#2{\setbox\lett=\hbox{#2}\lheight\ht\lett
\multiply\lheight by 12 \divide\lheight by 10\relax%
\lwidth\wd\lett \multiply\lwidth by 8 \divide\lwidth by 10\relax #2\kern-\lwidth%
\raise\lheight\hbox{{$\scriptstyle #1$}}\kern.1ex}

\begin{document}

\preprint{RUB-TPII-03/05}

\title{Solving the leading order evolution equation for GPDs }

\author{A. Manashov}
\altaffiliation[Also at ]{Department of Theoretical Physics,  Sankt-Petersburg State University,
St.-Petersburg, Russia}
\affiliation{Institut f\"ur Theoretische Physik, Universit\"at
                         Regensburg, D-93040 Regensburg, Germany}
\author{M. Kirch}
\affiliation{Institut f\"ur Theoretische Physik, Ruhr-Universit\"at Bochum,
                          D-44780 Bochum, Germany}
\author{A.~Sch\"afer}
\affiliation{Institut f\"ur Theoretische Physik, Universit\"at
                         Regensburg, D-93040 Regensburg, Germany}
\date{\today}

\begin{abstract}
An analytic method for the solution of the evolution equation for GPDs is
presented. The small $x,\xi$ asymptotics of GPDs are calculated.
\end{abstract}
\pacs{12.38.-t, 12.38.Bx}
\maketitle

One of the outstanding problems of quantum-chromodynamics is to
understand how hadrons are built from quarks and gluons and
how hadron properties result from their microscopic quark-gluon 
structure. 
In recent  years  
the formalism of generalized
parton distributions (GPDs) was developed~\cite{GPDs,Diehl},
which offers the most economic and comprehensive possibility to 
describe the internal structure of hadrons. GPDs
provide a general framework covering all relevant processes, from
inclusive ones (deep inelastic scattering) to 
exclusive ones,
and, therefore, allow us to combine the information contained in 
different measurements in an optimal manner. In addition, GPDs link this
information 
to fundamental elements of internal hadron
structure, which  cannot be directly determined by a measurement,
like information on the
total angular momentum of quarks or the spatial quark distribution in 
a fast moving hadron.

For instance, knowledge of the transverse spatial structure of quark and 
gluon distributions in a 
very high-energy proton is needed to describe proton-proton
collisions at the LHC (CERN) which will start operating next year.
The reason is that at LHC-energies several simultaneous hard interactions
will occure in a single proton-proton collison. The rates for such 
events depend crucially on how quarks and gluons of different momentum 
fractions $x$ 
are distributed in the transverse direction.
Without a quantitative understanding of  this QCD-background e.g.
the experimental search for physics beyond the 
standard model will be less sensitive. 

In practice, the determination of GPDs is highly
non-trivial, because they typically enter only in convolutions.
Probably only global fits to all relevant experimental \cite{GPDexp}
and lattice data \cite{GPDlatt} will be selective enough to really
determine more than the most dominant GPDs. Such fits naturally rely
heavily on $Q^2$ evolution. 
The evolution equations for GPDs are well known
~\cite{RB}; however, analytic methods to solve them were lacking.
As a result, one has to adhere to numerical methods which are
usually  
cumbersome and  often unstable. Obviously, this whole field
would profit enormously, if a faster and numerically more stable
method to calculate
the $Q^2$ evolution of GPDs could be found. In this contribution we
present a simple algorithm, defined by Eqs. (\ref{eq1}) - (\ref{final}) which 
fulfills these demands for LO evolution. 
This algorithm is based on an analytic method for the determination of 
the GPD scale dependence, which relies heavily on the symmetries of the
evolution equations. 

The analytic structure of GPD evolution
is intrinsically related to
that of  Dokshitzer-Gribov-Lipatov-Altarelli-Parisi (DGLAP) evolution \cite{DGLAP} for inclusive
processes and Efremov-Radyushkin-Brodsky-Lepage (ERBL) evolution \cite{ERBL} for exclusive processes,
because all three sets of evolution equations 
are only different facets of  one and the same equation --
the renormalization group equation for 
the  corresponding composite operators.
In this Letter we consider as concrete examples
the gluon GPDs of the  nucleon 
($H^g(x,\xi,t)$, $E^g(x,\xi,t)$) related to the
matrix element of the twist-two gluon operator
\begin{equation}\label{O12}
{\cal O}(z_1,z_2)=\tr\left\{G^{+\mu}(z_1 n)G^{+}_\mu(z_2n)\right\}\,,
\end{equation}
but the method is general. We also  neglect mixing with quark-antiquark
operators. (For the exact definitions of the GPDs and kinematical variables,
$x,\xi, P^+$ see  Ref.~\cite{Diehl}.) Since all of these GPDs  obey the same
evolution equation we introduce the common notation, $\varphi_\xi(x)$.

Because the derivation of our solution is rather technical let us
state already here the basic result. We find it crucial to treat the
two kinematic regions $|\xi|<|x|$ (the DGLAP region) and $|x|<|\xi|$
(the ERBL region) differently. Given some GPD parametrization at some
input scale $\mu_1$, one calculates the coefficients $c_{\xi}(j)$,
$c_{\xi}^\pm(j)$ from
\ba
\label{eq1}
c_\xi(j)&=&\frac{2j-1}{2}\int_{-1}^1 dx \,
\partial_x^2\,p_{j}({x})\,
\partial_x^2\,[\varphi_\xi^{\mu_1}(x|\xi|)]\,,
\\
\label{eq2}
c^{\pm}_\xi(j)&=&\frac{2j-1}{2}\int_1^{1/\vert\xi\vert} dx \,\partial_x^2\,p_{j}({x})\,
\partial_x^2\,[\varphi_\xi^{\mu_1}(\pm x|\xi|)]
\ea
and plugs these into \footnote{Since the GPD $\varphi_\xi(x)$ is an even function of 
$x$ and $\xi$ the expansion coefficients $c_\xi(j)$ vanish for even $j$, and the functions 
$c_\xi^\pm(j)$ are equal to each other, $c_\xi^+(j)=c_\xi^-(j)$}
\ba
\label{final}
\varphi_\xi^{\mu_2}(x)&=&
\Theta(|\xi|-|x|)\sum_{j=3}^{\infty}\,
c_\xi^{\mu_1}(j)\, L^{-\gamma(j)}\,
p_{j}\left(\frac{x}{|\xi|}\right)
\nonumber
\\
&+&
\sum_{a=\pm}
\frac{1}{\pi i}\,
\int_{C}dj\,
\,c^{\mu_1,a}_\xi(j)L^{-\gamma(j)}\, {q}_j
\left(\frac{ax}{|\xi|}\right).
\ea
Here
$L=\frac{\alpha_s(\mu_1)}{\alpha_s(\mu_2)}$,\
and the anomalous dimension $\gamma(j)$ is
specified later. 
The integration goes along the line parallel to the imaginary axis such that $2<\mathrm{Re}j<3$.
 The functions $p_j(x)$, $q_j(x)$ are expressed in terms of  the Legendre functions
of the first and second kinds~\cite{GR}
\ba\label{pj}
p_j(x)&=&
(1-x^2)\,\mathrm P_{j-1}^{-2}(x)\,,\\
q_j(x)&=&(1-x^2)
\frac{e^{i\pi j} \mathrm{Q}^{-2}_{j-1}(x_+)-
e^{-i\pi j}\mathrm{Q}^{-2}_{j-1}(x_-) }{2i\sin\pi j},
\ea
where $x_\pm=x\pm i0$~\footnote{
Note that the definition of Legendre function $\mathrm{ P}_{j-1}^{-2}(x)$
for $|x|>1$ and $|x|<1$ is different~\cite{GR}.}.
The sums and  integrals in (\ref{final}) converge absolutely.
As it is  clear from (\ref{final}) the first term contributes
only to the ERBL region,
while the integral gives contributions both to the DGLAP and ERBL regions.
We notice also that function $q_j(x)=0$ for $x<-1$,
while for  $x>1$ and  $|x|<1$ it can be simplified to
\ba\label{qq}
q_j(x)&=&(1-x^2)\mathrm{ Q}_{j-1}^{-2}(x),
\hskip 1 cm
x>1
\\
\label{qq-1}
q_j(x)&=&
-\frac{\pi}{2\sin\pi j}(1-x^2)\mathrm{ P}_{j-1}^{-2}(-x),
\hskip 0.5 cm
|x|<1\,.
\ea
It vanishes fast with $j\to \pm i\infty$.
The formula (\ref{final}) can be used both for numerical and analytic studies
of GPD evolution.
The main point is that if one chooses the GPD as a simple
analytic form at the starting value $\mu_1$, the integrals in (\ref{eq1}) and (\ref{eq2})
can be done analytically. Then to restore GPD at the scale $\mu_2$ one has
to evaluate the sum and the one-dimensional integral.
We also stress that the
form of Eq.~(\ref{final}) is completely determined by the symmetry properties
of the evolution equation.

Below we explain the main steps in the derivation of Eq. (\ref{final}), the details
will be given elsewhere.
The evolution equation for the GPD $\varphi_\xi(x)$ follows from that
for the  operator~(\ref{O12}). The corresponding kernel~(see
refs.~\cite{BFLK,RB})
is usually given in momentum space (i.e. as a function of $x$).
For our purposes it is more convenient
to use the coordinate-space formulation~\cite{BB}
\begin{eqnarray}\label{Ft}
\Phi_\xi(z)&=&(P^+)^{5/2}
\int_{-\infty}^\infty
 dx e^{iP^+xz} \varphi_\xi(x)\,,
\\
\varphi(z_1,z_2)&=&\frac{(P^+)^{1/2}}
{2\pi}\int_{-\infty}^\infty d\xi
 \,e^{-iP^+\xi(z_1+z_2)} \Phi_\xi(z_{12})\,,
\nonumber
\end{eqnarray}
where $z_1$ and $z_2$ are real variables, and $z_{12}=z_1-z_2$.
The LO evolution equation for $\varphi(z_1,z_2)$ reads
\be\label{RG}
\left(\mu\frac{\partial}{\partial\mu}+\beta(g)
\frac{\partial}{\partial g}\right) \varphi(z_1,z_2)=-
N_c\frac{\alpha_s}{\pi}\,[\mathbb{H}\,\varphi](z_1,z_2)\,.
\ee
The integral operator $\mathbb{H}$~\cite{BB} can be cast into the form
\begin{eqnarray}
[\mathbb{H}\,{\varphi}](z_1,z_2) & = & -4\int_0^1\!\! d\alpha\int_0^{\bar\alpha}\!d\beta
(\bar\alpha\bar\beta+2\alpha\beta){\varphi}( z_{12}^{\alpha},z_{21}^\beta)\nonumber\\
& & +\!\int_0^1 d\alpha\frac{\bar\alpha^2}{\alpha}
\left[2{\varphi}(z_1,z_2)-{\varphi}(z_{12}^\alpha,z_2)\right.\nonumber\\
& & \left.-{\varphi}(z_1,z_{21}^\alpha)\right]\!+\!\frac76{\varphi}(z_1,z_2).\label{HAM-1}
\end{eqnarray}
Here we use shorthand notations $z_{ik}^\alpha=z_i\bar\alpha+z_k\alpha$,
\mbox{$\bar\alpha=1-\alpha$}.
The Hamiltonian (\ref{HAM-1}) commutes
with the generators of collinear conformal transformations
$S^a=(S^a_1+S^a_2)$, $a=\pm,0$;
~$[\mathbb{H},\,S^{a}]=0$.
The one particle operators, $S^{a}_k$, are the generators
of the $SL(2,R)$  group
\be
S_k^-=-\partial_k,\ \ \ S^+_k=z_k^2\partial_k+2s_k z_k, \ \ \ S_k^0=z_k\partial_k+s_k\,,
\ee
in the representation  with spin  $s_k=3/2$.
The corresponding finite transformations $T(g)$ maps the function
$f(z)$
of the real variable $z$ onto another one according to
\be
[T(g^{-1})f](z)=\frac{1}{(cz+d)^3}\,
f\left(\frac{az+b}{cz+d}\right)\,,
\label{Tg}
\ee
where $a,b,c,d$ are the entries of the matrix $g\in SL(2,R)$.
To solve the  LO evolution equation this symmetry plays a crucial role.

The GPD $\varphi(z_1,z_2)$ depends on two real variables $z_1,z_2$ and transforms
under $SL(2,R)$ transformations according to  the tensor product of
two representations~(\ref{Tg}). Solving the evolution equation, one
seeks for the
complete system of functions that diagonalize the
Hamiltonian. Usually such a basis 
is provided by the eigenfunctions of the Casimir operator of the
symmetry group.
In the case under consideration the Casimir operator is
\be\label{Casimir}
\mathbb{J}^2=S^+S^-+S^0(S^0-1)=-z_{12}^{-1}\partial_1\partial_2\,z_{12}^3\,.
\ee

To solve the eigenvalue problem for the Casimir operator~(\ref{Casimir}) one should
specify the scalar product.  Our choice is restricted by the requirement of the
$SL(2,R)$ invariance of the latter and by the condition for  the ``physical''  GPD
to be normalizable with respect to this scalar product.
These requirements lead us to
\be\label{sc}
||\varphi||^2=\int dz_1dz_2\,z_{12}^4\,|\varphi(z_1,z_2)|^2=
\frac12\int d\xi dx\,
|\partial_x^2\varphi_\xi(x)|^2\,.
\ee
One can check that (\ref{sc}) is invariant with respect
to the transformations~(\ref{Tg}) and
that  the total spin operators  $S^{\pm,0}$
are antihermitean.
The Casimir operator~(\ref{Casimir})
is self-adjoint and  its
eigenfunctions form a basis in the Hilbert space with
the scalar product~(\ref{sc}).

The eigenfunctions of the Casimir operator can be easily found and they are in
accordance with the decomposition of the tensor product of two representations
(\ref{Tg}) into irreducible ones~\cite{Gelfand}. The operator
$\mathbb{J}^2$ has both a
discrete and continuous spectrum. The corresponding eigenfunctions can be expressed
as
\ba
\label{Ef-d}
P_{j=n+1}^{\xi}(z_1,z_2)&=&\sqrt{P^+}e^{-iP^+\xi(z_1+z_2)}\, \Psi_{j}^\xi(z_{12})\,,
\\
\label{Ef-c}
P_{j=\frac12+i\rho}^{\xi,\pm}(z_1,z_2)
&=&\sqrt{P^+}e^{-iP^+\xi(z_1+z_2)}\, \Psi_{j}^{\xi,\pm}(z_{12}),
\ea
where $n\in Z_+$ and $\rho\in R_+$.
The functions $\Psi_{j}^{\xi,\pm}(z)$ and $\Psi_{j}^{\xi}(z)$ are
defined for arbitrary complex $j$ as follows
\ba\label{Psi+}
\Psi_{j}^{\xi,\pm}(z)&=&
\frac1{2\cos\pi j}\left[\Psi_{j}^\xi(z_\pm)
-\Psi_{1-j}^\xi(z_\pm)\right ]\,, \\ \label{Psid}
\Psi_{j}^\xi(z)&=&
e^{-i\frac{\pi}{2}(j-1/2)}\,z^{-5/2} \,J_{j-1/2}\left(|P^+\xi|z\right)\,,
\ea
where $J_{\nu}(z)$ is the Bessel function and $z_\pm=\pm z+i0$.
(Let us note that by construction $\Psi_{j}^{\xi,\pm}(z)=
\Psi_{1-j}^{\xi,\pm}(z)$.) The eigenvalue of the Casimir operator in both
cases is $E=j(j-1)$. Taking the Fourier transform of (\ref{Ef-d}) one finds
 that the eigenfunctions
of the discrete and continuous spectrum have support in the
ERBL and DGLAP regions, respectively.

The decomposition of the arbitrary function $\varphi(z_1,z_2)$
into eigenfunctions
of the Casimir operator thus reads
\ba\label{dec-1}
\varphi(z_1,z_2)&=&
\int \frac{d\xi}{2\pi}
 \,
\left\{
\sum_{j=1}^{\infty}\,\omega(j)\, a_\xi(j)\,P_{j}^{\xi}(z_1,z_2)
\right. \\
&-&\left.i\int_{1/2-i\infty}^{1/2+i\infty} dj\, \omega^c(j)\,
a^\pm_\xi(j)\,P_{j}^{\xi,\pm}(z_1,z_2)
\right\},\nonumber
\ea
where $\omega(j)=2j-1$ and $\omega^c(j)=(j-1/2)\cot\pi j$. The expansion coefficients
$a_\xi(j)$ and $a_\xi^\pm(j)$
are given by the scalar product of  $\varphi(z_1,z_2)$ with the eigenfunctions
(\ref{Ef-d}) and (\ref{Ef-c}).
We get rid of the integral over $\xi$ and write down the expansion directly
for the function $\Phi_\xi(z)$ (\ref{Ft}) as follows
\ba
\label{dec-2}
\Phi_\xi(z)&=&
\sum_{j=3}^{\infty}\,(2j-1)\, a_\xi(j)\,\Psi_{j}^{\xi}(z)
\\
&&-
\frac{i}2\int_{5/2-i\infty}^{5/2+i\infty} \frac{dj}{\sin\pi j}\, (2j-1)\,
a^\pm_\xi(j)\,\Psi_{j}^{\xi}( z_\pm) \,.\nonumber
\ea
To derive (\ref{dec-2}) we made use of  the explicit form of the functions (\ref{Psi+})
and the symmetry,  $a_\xi^\pm(j)=a_\xi^\pm(1-j)$, of the expansion coefficients.
We notice also that the difference between the integrals in (\ref{dec-1}) and (\ref{dec-2})
can be calculated by residues and reproduces  the missing terms ($j=1,2$) in the
sum in the Eq.~(\ref{dec-2}).

Now we are ready to discuss the evolution of the function
$\Phi_\xi(z)$ driven by the
Hamiltonian~(\ref{HAM-1}). The eigenfunctions of the discrete spectrum
of the  operator  $\mathbb{J}^2$ are the eigenfunctions of the
Hamiltonian
with energies
$$
E(j)=2\left[\psi(j)-\psi(3)-\frac{1}{j(j+1)}-\frac{1}{(j-1)(j-2)}\right]+\frac76.
$$
Let us notice that $E(j)$  is defined only for integer $j\geq3$.
Further, one can check that the eigenfunctions of the continuous spectrum~(\ref{Ef-c})
 do not
diagonalize  $\mathbb{H}$. This seems to contradict the
commutativity of  $\mathbb{H}$
and $\mathbb{J}^2$.
 The explanation is simple,
the Hamiltonian $\mathbb{H}$ is not self-adjoint
with respect to the scalar product~(\ref{sc}) and thus does not have the
same eigenfunctions as the operator~$\mathbb{J}^2$~\footnote{It means that
the Hamiltonian $\mathbb{H}$ does not commute with finite $SL(2,R)$ transformations.}.

Nevertheless, one can find functions which diagonalize the Hamiltonian (of course, they are
not mutually orthogonal and do not form a basis of the Hilbert space).
Indeed, the  equation $\mathbb{J}^2\Psi(z_1,z_2)=j(j-1)\Psi(z_1,z_2)$ (here $j$ is
an arbitrary
complex number) after
separation of the factor
$e^{-iP^+\xi(z_1+z_2)}$,
turns into a second order differential equation with the
two independent solutions
$\Psi^\xi_j(z)$ and  $\Psi^\xi_{1-j}(z)$.
Because of the commutativity of the integral operator
$\mathbb{H}$ and the differential operator $\mathbb{J}^2$,
one concludes that
\be\label{HH}
\mathbb{H}\Psi^\xi_j(z)=A(j)\Psi^\xi_j(z)+B(j)\Psi^\xi_{1-j}(z)\,.
\ee
Substituting $\varphi(z_1,z_2)=e^{-iP^+\xi(z_1+z_2)}\Psi^\xi_j(z)$
into  (\ref{HAM-1}) one finds that the integrals converges
when $\mathrm{Re}j>2$.
Next, let us notice that when
the variables $\alpha$ and $\beta$  run over  the integration region
the argument of the function $\Psi^\xi_j$ varies from $0$ to $z$.
Thus to fix  the coefficients $A(j)$ and $B(j)$ it is sufficient to study
the $z\to 0$ asymptotics of the r.h.s. and l.h.s. of (\ref{HH}). In this case one can
substitute
$\Psi^\xi_j(z)$ by its leading term
$\sim z^{j-3}$ \  to get
$
A(j)=E(j),\ B(j)=0\,.
$
Thus we have shown  that the function $e^{-iP^+\xi(z_1+z_2)}\Psi^\xi_j(z)$ diagonalizes the
Hamiltonian $\mathbb{H}$.

Now we are in the position to solve  the evolution equation for
the GPD $\Phi_\xi(z)$. Indeed,
the decomposition (\ref{dec-2}) involves only
the functions $\Psi^\xi_j(z)$ which diagonalize the Hamiltonian, and
the integration follows the  line $\mathrm{Re}\,j=5/2$.
So we can apply the Hamiltonian
$\mathbb{H}$, Eq.~(\ref{HAM-1}),  and change the order of
integration between  $j$
and $\alpha,\beta$.
Thus the solution of the evolution equation can be written in the
form~(cf. \cite{BB, KM})
\ba
\label{evol-sol}
\Phi_\xi^{\mu_2}(z)&=&
\sum_{j=3}^{\infty}\,(2j-1)\, a_\xi^{\mu_1}(j)\, L^{-\gamma(j)}\,\Psi_{j}^{\xi}(z)
\\
 &-&
\frac{i}{2}\int_C \frac{dj}{\sin\pi j}\, (2j-1)\,
a^{\mu_1,\pm}_\xi(j)\,L^{-\gamma(j)}\,\Psi_{j}^{\xi}(z_\pm)\,.\nonumber
\ea
Here
$\gamma(j)=2N_c\,E(j)/b_0$, $ b_0=\frac{11}{3}N_c-\frac23 n_f$.
After  Fourier transformation,  Eq.~(\ref{evol-sol}) can be cast
into the form~(\ref{final}).
\vskip 0.2cm

Let us discuss the properties of the  solution (\ref{final}).
We represent GPD $\varphi_\xi$ as  $\varphi_\xi(x)=\varphi_\xi^{I}(x)+\varphi_\xi^{II}(x)$,
where $\varphi_\xi^{I}(x)$ is given by the sum in Eq. (\ref{final}) and
$\varphi_\xi^{II}(x)$ by the integrals.
The function $\varphi_\xi^{I}(x)=0$  outside the ERBL region.
The function  $\varphi_\xi^{II}(x)=\varphi_\xi(x)$ in the DGLAP region
and does not vanish for $|x|<|\xi|$. At the scale $\mu_1$
it can be calculated  for $|x|<|\xi|$ in closed form,
$\varphi_\xi^{II}(x)=\varphi_\xi(\xi)+\varphi'_\xi(\xi) (x^2-\xi^2)/2\xi$.
This contribution is entirely due to the  eigenfunctions of the discrete spectrum
with $j=1,2$ (see Eqs. (\ref{dec-1}), (\ref{dec-2})).

As it was stated earlier the sum and the integrals in (\ref{final}) are absolutely convergent.
Thus having calculated the coefficients $c_\xi^\mu(j)$ and  $c_\xi^{\mu,\pm}(j)$ at scale
$\mu_1$ one can easily restore the function at another scale $\mu_2$.

Let us also note that the evolution of the coefficients $c_\xi(j)$ and $c_\xi^\pm(j)$ is
not autonomous. To find
the  coefficients at scale $\mu_2$ one should insert the function $\varphi_\xi^{\mu_2}(x)$
into the Eqs.~(\ref{eq1}) and (\ref{eq2}). Since the function $q_j(x)$ is not an eigenfunction
of the Casimir operator one easily figures out, e.g. that the coefficient $c_\xi^{\mu_2}(j)$
cannot be expressed solely in terms of the coefficient $c_\xi^{\mu_1}(j)$. Rather,
after some algebra, one finds
\ba\label{c-n}
c_\xi^{\mu_2}(j')&=&c_\xi^{\mu_1}(j')
L^{-\gamma(j')}+
\int_C \frac{dj}{2\pi i} \frac{1}{(j'+j-1)(j-j')}\nonumber\\
&\times&L^{-\gamma(j)}\left[c_\xi^{\mu_1,+}(j)+(-1)^{j'-1}c_\xi^{\mu_1,-}(j)\right]\,.
\ea

The basic Eq.~(\ref{final})  can be analyzed analytically
in different limits. As an example let us discuss the asymptotic expansion of the GPD
at large $\mu_2$ (see Refs.~\cite{RB,BGMS}).
To this end we shift the integration contour from the line
$\mathrm{Re}\ j=5/2$ to the line $\mathrm{Re}\ j=N-1/2$,  $N$ being an integer.  It follows
from the Eqs.~(\ref{qq}) and 
(\ref{qq-1})  that for $x>1$ the function $q_j(x)$ is analytic for $\mathrm{Re}\ j>2$,
while for $|x|<1$ it has poles at integer $j$. Thus, shifting the contour one has to evaluate
residues at these points. The integral over the new contour decreases faster than
$L^{-\gamma(N-1/2)}$. Keeping the first $N-3$ terms from the sum and adding the contribution
from the residues one finds
\ba\label{as}
\varphi_\xi^{\mu_2}(x)&=&\Theta(|\xi|-|x|)\sum_{j=3}^{N-1}
D_\xi^{\mu_1}(j)\,L^{-\gamma(j)}\\
&&\times\left(1-\frac{x^2}{\xi^2}\right)^2\,C_{j-3}^{5/2}
\left(\frac{x}{|\xi|}\right)
+{\mathcal O}(L^{-\gamma(N)})\,.\nonumber
\ea
Here $C_{j-3}^{5/2}$ is the Gegenbauer polynomial and the coefficient $D_\xi^{\mu_1}(j)$ reads
\be\label{D}
D_\xi^{\mu_1}(j)=r_j\int_{-1/|\xi|}^{1/|\xi|}dx\,
C_{j-3}^{5/2}(x)\,\varphi_\xi^{\mu_1}(x|\xi|)\,,
\ee
where $r_j={9/2(2j-1)}{\Gamma(j-2)}/{\Gamma(j+2)}$.

\begin{figure}[t]
\psfrag{a}[cc][rc]{$a=x/\xi$}
\psfrag{r}[cc][cc]{$R(a)$}
\centerline{\epsfxsize6.0cm\epsfbox{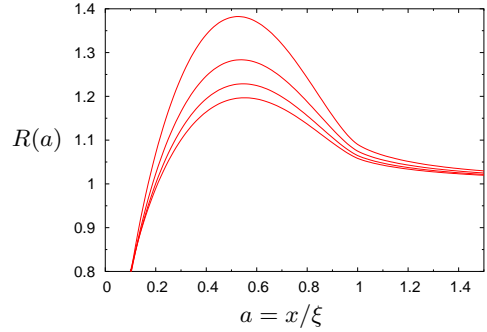}}
\vspace*{0.5cm}
\caption[]{The ratio $R(a)=R^+(a)+R^+(-a)$
for $L=2.4$ and $x=10^{-3}$ (upper curve), $10^{-4}$, $10^{-5}$
and $10^{-6}$ (lowest curve).}
\label{GPD}
\end{figure}

As another illustration let us  study  the small $\xi$ behavior
of GPDs.
For small $\xi$ (we take $\xi>0$) the coefficient $c^+_\xi(j)$
can be represented in the form
\be\label{cxi}
c^{\mu_1,+}_\xi(j)=f(j) \bar c^{\mu_1,+}(j)
({2}/{\xi})^{j-1}-(j\to 1-j)\,,
\ee
where $f(j)=-\frac{\xi}{\sqrt{\pi}}{\Gamma(j+1/2)}/{\Gamma(j-2)}$ and the coefficient
$\bar c^{\mu_1,+}(j)=\int_0^{1}dx\,x^{j-3} \varphi_{\xi=0}^{\mu_1}(x)+{\mathcal O}(\xi^2)$. 
Denoting by $\varphi_{\xi}^{\mu_2,+}(x)$ the integral corresponding to the term with $a=+$ 
in the sum (\ref{final}) and taking into account (\ref{cxi}) one finds for $x\gg\xi$ 
\be\label{xggxi}
\varphi_\xi^{\mu_2,+}(x)=\frac{1}{2\pi i}\int_{C} dj\,x^{-j+2}\,\bar c^{\mu_1,+}(j)\,
L^{-\gamma(j)}
+ {\mathcal O}(\xi^2)\,,
\ee
which corresponds to DGLAP evolution.
It is interesting to compare the asymptotics in two regimes: $\xi\to 0$, $x$ fixed 
and $\xi\to 0$, $x=a\xi$. Using (\ref{xggxi}) and (\ref{final}) one derives 
for the ratio of the asymptotics
\be\label{ration}
R^+(a)=
\frac{\varphi^{\mu_2,+}_{x/a}(x)}{\varphi^{\mu_2}_{0}(x)}=
\frac{e^{i\pi \nu} f_\nu(a_+)-e^{-i\pi \nu} f_\nu(a_-)}{2i\sin\pi\nu}\,,
\ee
where $a_\pm=a\pm i0$,  $\nu=\left(\frac{4N_c}{b_0}\frac{\log L}{\log 1/x}\right)^{1/2}$
and  
$$
f_\nu(a)=e^{-i\pi\nu\arg(a)}{}_2F_1(\nu/2,\nu/2+1/2,\nu+5/2|a^{-2}).
$$
Taking into account the similar contribution from the $a=-$ intergral in (\ref{final}) one obtains 
for the total ratio $R(a)=\varphi_{x/a}^{\mu_2}(x)/\varphi_{0}^{\mu_2}(x)$ 
the result shown in Fig.~\ref{GPD}.

To summarize, we have developed an analytic method
for solving the LO evolution equations for
GPDs. This method is quite general and relies on
the conformal symmetry  of the evolution equation.
Our approach can also be used to elucidate the analytic structure of three-particle
parton distributions which possess additional hidden symmetry~\cite{int}.

\begin{acknowledgments}
The authors are grateful to V.~Braun and D.~M\"uller for useful discussions.
This work was supported in part by the grant 03-01-00837
of the Russian Foundation for Fundamental Research (A.M.),
by the  Helmholtz Association (A.M. and A.S., contract number
VH-NG-004), and the Graduiertenkolleg 841 of DFG (M.K.).
\end{acknowledgments}


\end{document}